\documentclass[11pt,aps,nofootinbib,
floatfix,superscriptaddress]{revtex4}
\usepackage{graphicx}
\usepackage[mathscr]{eucal}
\usepackage{hyperref}
\begin{document}

\title{Global shift symmetry and vacuum energy of matter fields}

\begin{abstract}
{We construct the model incorporating both an arbitrary shift of cosmological
constant and Goldstone boson corresponding to the spontaneous breaking down
the global shift symmetry in the matter action. The gravity breaks down the
symmetry explicitly and transforms the Goldstone boson to the inflaton
field.}
\end{abstract}

\author{Ja.V.Balitsky,}
\affiliation{State Research Center of the Russian Federation  ``Institute for
High Energy Physics'' of National Research Centre  ``Kurchatov Institute'',
Russia, 142281, Moscow Region, Protvino, Nauki 1}

\author{V.V.Kiselev}
\email{Valery.Kiselev@ihep.ru} \affiliation{State Research Center of the
Russian Federation  ``Institute for High Energy Physics'' of National
Research Centre  ``Kurchatov Institute'',
Russia, 142281, Moscow Region, Protvino, Nauki 1}
\affiliation{Moscow
Institute of Physics and Technology (State University), Russia, 141701,
Moscow Region, Dolgoprudny, Institutsky 9}

\maketitle

\section{Global shift of vacuum energy density}
The cosmological constant of general relativity corresponds to the energy
density of vacuum  \cite{Weinberg:1988cp,Weinberg:2000yb}.  Let us show that
the cosmological constant itself is inherently related to a specific dynamics of matter fields.

Since forces of nature, except the gravity, do not depend on the value of
vacuum energy density $\rho_\Lambda$, equations of motion for fields of
matter are invariant under the global shift
\begin{equation}\label{rho-shift}
    \rho_\Lambda=\Lambda_0^4\mapsto \rho_\Lambda+\Lambda_0^4 \, u,\qquad
    \partial_\mu u\equiv 0.
\end{equation}
In this respect the cosmological constant in the action of matter fields
looks like an additional global scalar field.

If the action of matter fields is invariant under the global shift with the
parameter $u$ in (\ref{rho-shift}), then fixing the vacuum energy means the
spontaneous breaking down the global symmetry that leads to the appearance of
Goldstone--Nambu boson $\phi$ with an action invariant under the global shift
symmetry
\begin{equation}\label{phi-shift}
    \phi\mapsto\phi+f_G\,u,
\end{equation}
wherein $f_G$ is the constant of Goldstone boson.

The gravity is not invariant under the global shifts (\ref{rho-shift}) and
(\ref{phi-shift}) because the gravitational force depends on the absolute
value of energy. Therefore, the gravity breaks down the global invariance of
matter action. This breaking leads to generating an effective mass for the
boson $\phi$, instead of nil potential without the gravity, that makes it the
pseudo-Goldstone boson.

Such the program of induced pseudo-Goldstone scalar has been recently
considered in \cite{Balitsky:2014csa}, wherein we have argued for a
non-minimal interaction of $\phi$ with the curvature scalar and calculated
the effective potential in the one-loop approximation for the graviton
contribution. A cut-off in loop calculations is related with the Planck mass
and the coupling constant of non-minimal interaction, so that after a
conformal transformation we arrive to the gravity with the inflaton field
consistent with phenomenological parameters of early Universe inflation
\cite{Starobinsky:1980te,Mukhanov:1981xt,Guth:1980zm,Albrecht:1982wi,
Linde:1981mu,Linde:1983gd} and \cite{Agashe:2014kda}.

However, in such the treatment the general assumption of global invariance in
the matter field action has been suggested implicitly. The question is
whether the action with the global shift of matter fields and global shift of
vacuum energy density can be constructed explicitly.

\begin{figure}[t]
  \includegraphics[width=8cm]{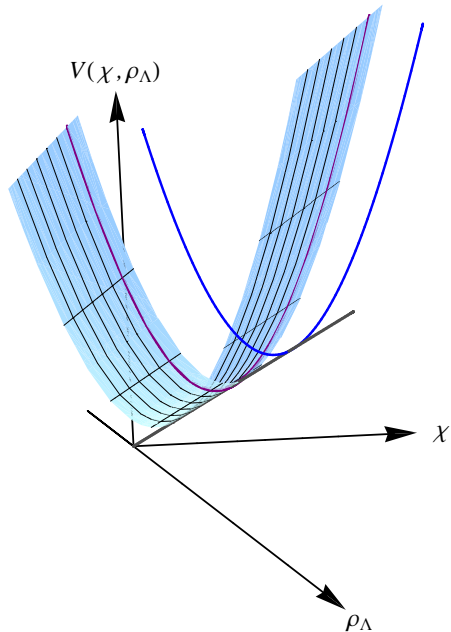}\\
  \caption{A potential of field $\chi$ versus the density of vacuum energy
  $\rho_\Lambda$.}\label{lin}
\end{figure}

This problem is not trivial, indeed. If the vacuum energy can take an
arbitrary value for the matter fields (without the gravity), then we have to
expect that a potential of matter fields is not restricted from the bottom.
This fact is illustrated in Fig. \ref{lin}, wherein an ordinary potential of
matter field $\chi$ depends on the density of vacuum energy $\rho_\Lambda$.
Then, the instability can occur.

In Section \ref{II} we present the model possessing the global shift symmetry
with the required properties. In Section \ref{III} we address the problem of
stability in two aspects: the limit of zero parameter of instability and
stochastic treatment for a source of field that could result in zero tensor
of energy-momentum for the stochastically averaged field. Section \ref{iV} is
devoted to short remarks about the Galilean symmetry
\cite{Deffayet:2009wt,Chow:2009fm,VanAcoleyen:2011mj,Nicolis:2008in} in the
model. In Conclusion we summarize the results and discuss its implementations
into the theory of inflaton field.

\section{Model\label{II}}
In the simplest and evident way, the action of two real scalar fields
$\psi_{1,2}$
\begin{equation}\label{act-12}
    S=\int\mathrm{d}^4x\,\mathscr L(\psi_1,\psi_2)
\end{equation}
possesses the global invariance
\begin{equation}\label{globe1}
    \psi_{1,2}\mapsto\psi_{1,2}+\sqrt{2}\,f_G\,u,
\end{equation}
if the Lagrangian is equal to
\begin{equation}\label{Lag}
    \mathscr L(\psi_1,\psi_2)=\frac12\left\{(\partial_\mu\psi_1)^2+
    (\partial_\mu\psi_2)^2
    \right\}+\frac{1}{\sqrt{2}}\,\Lambda_0^3(\psi_1-\psi_2).
\end{equation}
The transformation
\begin{equation}\label{trans}
    \phi=\frac1{\sqrt{2}}\,(\psi_1+\psi_2),\qquad
    \tilde\phi=\frac1{\sqrt{2}}\,(\psi_1-\psi_2),
\end{equation}
leads to the Lagrangian
\begin{equation}\label{Lag2}
    \mathscr L(\phi,\tilde\phi)=\frac12\left\{(\partial_\mu\phi)^2+
    (\partial_\mu\tilde\phi)^2
    \right\}+\Lambda_0^3\tilde\phi.
\end{equation}
The solutions of field equations for the vacuum state get the form
\begin{equation}\label{vac-sol}
    \langle\phi\rangle =v,\qquad
    \langle\tilde\phi\rangle=\tilde v+\frac12\,K_{\mu\nu}(x-x_0)^\mu(x-x_0)^\nu,
\end{equation}
with a symmetric tensor keeping the trace equal to
\begin{equation}\label{trace}
    K_\mu^\mu=\Lambda_0^3.
\end{equation}
At $K_{\mu\nu}x_0^\mu x_0^\nu=0$ or $x_0=0$, the contribution due to
$K_{\mu\nu}$ is irrelevant to the problem of constant
density of vacuum energy, while it refers to the instability of the model,
which will be treated in the next Section.
For the sake of simplicity we adopt the condition $x_0=0$, although one could
easily reformulate all of further statements for $x_0\neq0$, of course.

Then, we see that the field $\phi$
is the Goldstone boson possessing zero mass as well as the nil potential at
all, so that the action is invariant under the global shift
\begin{equation}\label{phi-globe2}
    \phi\mapsto\phi+f_G\,u,
\end{equation}
while $\tilde \phi$ is responsible for the global shift of cosmological
constant: $\rho_\Lambda=-\Lambda_0^3\tilde v$. The global shift of initial
fields generates the fractional shifts of vacuum energy density:
$$
    \psi_{1,2}\mapsto\psi_{1,2}+\sqrt{2}\,f_G\,u\quad
    \Rightarrow\quad (\delta\rho_\Lambda)_{1,2}=\mp\Lambda_0^3f_G\,u.
$$
The constant terms of expectations
$$
    \langle \psi_1\rangle_c=\frac1{\sqrt{2}}\,(v+\tilde v),\qquad
    \langle \psi_2\rangle_c=\frac1{\sqrt{2}}\,(v-\tilde v),
$$
spontaneously break down the global shift symmetry (\ref{globe1}).

Thus, the model explicitly generates the Goldstone boson under the
spontaneous breaking down the global shift symmetry.

\section{Erasing the instability\label{III}}
Since the potential in (\ref{Lag2}) is linear in $\tilde \phi$, it generates
the ``accelerated'' solution:
\begin{equation}\label{K1}
	\tilde\phi_A=\frac12\,K_{\mu\nu}x^\mu x^\nu,\qquad K_\mu^\mu=\Lambda_0^3.
\end{equation}
We treat such the solution as the instability, since it appears in the
vacuum, too. Let us discuss ways to cancel this instability under the
conservation of global symmetry properties and its spontaneous breaking.

\subsection{Limit of zero instability}
The simplest method is to take the limit of $\Lambda_0\to 0$ at
$$
	\Lambda_0^3\tilde v =\mbox{const.}
$$
Therefore,
\begin{equation}
	\tilde v=-\frac{\rho_\Lambda}{\Lambda_0^3}\to\infty.
\end{equation}
Thus, the instability is removed, while the cosmological constant could get
an arbitrary value $\rho_\Lambda$. In this scheme, the Goldstone boson can
take an arbitrary expectation value, of course. However, we arrive to the
couple massless fields.

\subsection{Stochastic source}
Consider the energy-momentum tensor
\begin{equation}\label{T1}
	T_{\mu\nu}=\partial_\mu\tilde \phi\,\partial_\nu\tilde \phi-g_{\mu\nu}\mathscr L
\end{equation}
for the ``accelerated'' solution in (\ref{K1}),
\begin{equation}\label{T2}
	T_{\mu\nu}\Big|_{\tilde\phi_A}=K_{\mu\nu'}x^{\nu'}\,K_{\nu\nu''}x^{\nu''}-
    \frac12\,g_{\mu\nu}\left\{g^{\mu'\mu''}K_{\mu'\nu'}x^{\nu'}\,K_{\mu''\nu''}x^{\nu''}+
    {\Lambda_0^3}\,K_{\mu'\nu'}x^{\mu'}x^{\nu'}
    \right\}.
\end{equation}
Since
$$
    K_\mu^\mu=\Lambda_0^3,
$$
the tensor of $K_{\mu\nu}$ represents the source for both the linear
potential and solution of field equations. Let us make this source to be
stochastic, i.e. the observable quantities are given by average values
obtained under the correlations of stochastic source. So, we immediately get
\begin{equation}\label{av1}
    \langle K_\mu^\mu\rangle=\Lambda_0^3.
\end{equation}
Further, in order to keep the Lorentz invariance we have to put
\begin{equation}\label{av2}
    \langle K_{\mu\nu}\rangle=\frac14\,g_{\mu\nu}\,\langle K\rangle,\qquad
    \langle K\rangle =\langle K_\mu^\mu\rangle=\Lambda_0^3,
\end{equation}
and analogously
\begin{equation}\label{av3}
    \langle K_{\mu\nu}\,K^\mu_{\nu'}\rangle=\frac14\,
    g_{\nu\nu'}\,\langle K^2\rangle,\qquad
    \langle K^2\rangle=\langle K_{\mu\nu}\,K^{\mu\nu}\rangle.
\end{equation}
Next, we fix the correlator
\begin{equation}\label{av4}
    \langle K_{\mu\nu}\,K_{\mu'\nu'}\rangle=A\left\{
    g_{\mu\mu'} g_{\nu\nu'}+g_{\mu\nu'} g_{\nu\mu'}\right\}+
    B\,g_{\mu\nu} g_{\mu'\nu'},
\end{equation}
so that
$$
    5A+B=\frac14\,\langle K^2\rangle,
$$
because
$$
    g^{\mu\mu'}\langle K_{\mu\nu}\,K_{\mu'\nu'}\rangle=\frac14\,g_{\nu\nu'}\,
    \langle K^2\rangle.
$$

Then, we can calculate the average value of energy-momentum tensor for the
``accelerated'' solution,
\begin{equation}\label{av5}
    \left\langle T_{\mu\nu}\Big|_{\tilde\phi_A}\right\rangle=
    g_{\mu\nu}\,x^2\,\left\{A-\frac18\,\left(\langle K^2\rangle+
    \langle K\rangle\,\Lambda_0^3\right)
    \right\}+
    x_\mu x_\nu\,(A+B).
\end{equation}
Tensor (\ref{av5}) becomes relativistically invariant if
\begin{equation}\label{AB}
    A+B=0,
\end{equation}
hence,
\begin{equation}\label{A-term}
    A=\frac{1}{16}\,\langle K^2\rangle,
\end{equation}
and
\begin{equation}\label{av6}
    \left\langle T_{\mu\nu}\Big|_{\tilde\phi_A}\right\rangle\Bigg|_\mathrm{inv}=
    -\frac{1}{16}\,g_{\mu\nu}\,x^2\,\left\{\langle K^2\rangle+2
    \Lambda_0^6    \right\}.
\end{equation}
The translational invariance of average tensor (\ref{av6}) takes place if
\begin{equation}\label{translat}
    \langle K^2\rangle=-2 \Lambda_0^6,
\end{equation}
so that
$$
    \left\langle T_{\mu\nu}\Big|_{\tilde\phi_A}\right\rangle
    \Bigg|_\mathrm{inv}\equiv 0.
$$
Thus, for the stochastic source with the correlators
\begin{equation}\label{corr-x}
    \langle K_{\mu\nu}\,K_{\mu'\nu'}\rangle=-\frac18\,\left\{
    g_{\mu\mu'} g_{\nu\nu'}+g_{\mu\nu'} g_{\nu\mu'}-
    g_{\mu\nu} g_{\mu'\nu'}\right\}\,\Lambda_0^6,\qquad
    \langle K_{\mu\nu}\rangle=\frac14\,g_{\mu\nu}\,\Lambda_0^3,
\end{equation}
the instability of ``accelerated'' solution completely disappears, since the
energy-momentum tensor of such the solution is stochastically equal to zero.

Note that the polarization operator
$$
    \Pi_{\mu\nu,\mu'\nu'}=\frac12\,\left\{
    g_{\mu\mu'} g_{\nu\nu'}+g_{\mu\nu'} g_{\nu\mu'}-
    g_{\mu\nu} g_{\mu'\nu'}\right\}
$$
corresponds to the projection of symmetric tensor $R^{\mu'\nu'}$ to its
Einstein partner,
$$
    G_{\mu\nu}=\Pi_{\mu\nu,\mu'\nu'}\,R^{\mu'\nu'}=R_{\mu\nu}-\frac12\,R
    \,g_{\mu\nu}.
$$

Finally, the  shift of ``accelerated'' solution
$$
    \tilde\phi_A\mapsto \tilde \phi_V=\frac12\,K_{\mu\nu}x^\mu x^\nu+\tilde v
$$
and the stochastic averaging give the vacuum with the energy-momentum tensor
equal to
$$
    \langle T_{\mu\nu}\rangle=-\Lambda_0^3\tilde v\,g_{\mu\nu},
$$
as we have expected.

\section{Galilean invariance\label{iV}}
In Minkowski space-time, equations for the motion of both fields $\phi$ and
$\tilde\phi$  possess the invariance under the Galilean transformations
\cite{Deffayet:2009wt,Chow:2009fm,VanAcoleyen:2011mj,Nicolis:2008in}
\begin{equation}\label{Gal1}
    \phi\mapsto \phi+b_\mu x^\mu,\qquad
    \tilde\phi\mapsto \tilde\phi+\tilde b_\mu x^\mu.
\end{equation}
The stochastic averaging for $\tilde \phi$ reduces the Galilean invariance to
the global shift of the field, if we consider the energy-momentum tensor,
when such the field transformation is equivalent to the introduction of
coordinate translation in the form of ``accelerated'' solution with an
appropriate introduction of change in the constant term of the field.

The Galilean invariance of Goldstone boson remains essential especially in
the procedure of implementation of invariant interactions of Goldstone boson
with itself, matter and gravity
\mbox{\cite{Deffayet:2009wt,Chow:2009fm,VanAcoleyen:2011mj,Nicolis:2008in}.}

\section{Conclusion}
We have just shown that the invariance of non-gravitational forces with
respect to the variation of cosmological constant corresponds to the specific
dynamical symmetry of global shift in the action of matter fields, while the
spontaneous breaking down this global symmetry leads to the appearance of
Goldstone boson with the additional Galilean symmetry of its interactions. In
order to provide the stable theory we have offered the mechanism of
stochastic source, which guarantees the Poincare-invariant value of
energy-momentum tensor for the scalar field being under the danger of
instability.

The gravitation breaks down the global invariance explicitly, hence, the
Goldstone boson acquires an effective potential due to the interaction with
the gravity, i.e. due to the graviton loops. The one-loop approximation gives
the model of inflaton \cite{Balitsky:2014csa}, which is consistent with the
modern observations \cite{Agashe:2014kda}. Moreover, such the origin of
inflaton can drive the mechanism for suppressing the cosmological constant
\cite{Balitsky:2014epa}.

In this respect we have to mention other approaches to both problems of
inflaton origin and cosmological constant with various motivations.

So, the pseudo-Goldstone nature of inflaton field is used in \emph{i)}
supersymmetric models with flat directions in K\"ahler potential
\cite{Kawasaki:2000yn,Kallosh:2010ug}, \emph{ii)} the ``natural inflation''
with the axion \cite{Freese:1990rb}, \emph{iii)} the induced gravity with the
scale invariance \cite{Csaki:2014bua}.

The theories with a non-metric measure of volume
\cite{Guendelman:1995xe,Guendelman:1996jr,Guendelman:1999jb,
Guendelman:1999qt,Guendelman:1999rj,Guendelman:1999tb,Guendelman:2001xy,
Guendelman:2009ck,Guendelman:2014bga,Guendelman:2015rea} operate with the
global shift of cosmological constant and give models which are consistent
with switching the inflaton potential from the huge plateau in the early
Universe into a suppressed plateau at modern times.

Let us comment on the theory of non-metric measure, wherein one makes using
the substitution $\sqrt{-\det g_{\mu\nu}}\mapsto
\epsilon^{\mu\nu\alpha\beta}\partial_\mu\mathcal A_{\nu\alpha\beta}$. So, in
the Minkowski limit of $\sqrt{-\det g_{\mu\nu}}\mapsto 1$, we get the action
in the form
$$
    S=\int\mathrm{d}^4x\,\mathcal L(\Phi,\partial \Phi)\,\partial_\mu A^\mu,
$$
where $\Phi$ is a field of matter,
$A^\mu=\epsilon^{\mu\nu\alpha\beta}\mathcal A_{\nu\alpha\beta}$ is the field
density dual to the initial field of $\mathcal A_{\nu\alpha\beta}$. Then, the
Lagrange--Euler equations for $A^\mu$ read off
$$
    \partial_\mu L\equiv 0,
$$
that means
$$
    L=\Lambda^4=\mbox{const.}
$$
Therefore, denoting $\partial_\mu A^\mu=\tilde \phi/\Lambda$ we arrive to the
action with the linear potential of field $\tilde\phi$,
$$
    S=\int\mathrm{d}^4x\,\Lambda^3\tilde\phi,
$$
in the form very analogous to our consideration. In
\cite{Guendelman:1995xe,Guendelman:1996jr,Guendelman:1999jb,
Guendelman:1999qt,Guendelman:1999rj,Guendelman:1999tb,Guendelman:2001xy,
Guendelman:2009ck,Guendelman:2014bga,Guendelman:2015rea} the field
$\tilde\phi$ is nontrivially related with the matter and gravitational
fields, that constitutes the difference with the simple approach offered in
the present paper. By the way, we also note that a non-metric volume measure
with density $\Psi$ is equivalent to the introduction of specific inteaction
with the metric, since $\Psi=U(\Psi,g_{\alpha\beta})\cdot\sqrt{-\det
g_{\mu\nu}}$ at the potential $U=\Psi/\sqrt{-\det g_{\mu\nu}}$. To the
current moment, the origin of such the interaction lies beyond any
argumentation in the framework of symmetries like the gauge invariance, to
our opinion, although the global scale invariance is the favorite of such the
motivation mentioned above.

Next, the idea of dynamical evolution is applied to the cosmological constant
in \cite{Shapiro:2000dz,Shapiro:2001rh,Sola:2007sv,Shapiro:2009dh,
Sola:2013gha,EspanaBonet:2003vk,Gomez-Valent:2014rxa} that transform the
problem of cosmological constant to the problem of dark energy.

Thus, we see that the approach offered in the present paper can be used to
differentiate complex models of particle physics versus their relevance to
the inflaton physics and cosmological constant problem: the most prospective
models of matter fields should involve the global invariance spontaneously
broken by setting the vacuum expectation values and arbitrary cosmological
constant, while the interaction with the gravity should break down the global
invariance explicitly.

This work is supported by Russian Foundation for Basic Research, grant \#
15-02-03244.

\bibliography{bibglobe}

\end{document}